# catwoman: A transit modelling Python package for asymmetric light curves

Kathryn Jones[1] and Néstor Espinoza[2]

**1** University of Bern, Center for Space and Habitability, Gesellschaftsstrasse 6, CH-3012, Bern, Switzerland **2** Space Telescope Science Institute, 3700 San Martin Drive, Baltimore, MD 21218, USA

## Summary

When exoplanets pass in front of their stars from our point of view on Earth, they imprint a transit signature on the stellar light curve which, to date, has been assumed to be symmetric in time, owing to the planet being modelled as a circular area occulting the stellar surface (Kreidberg, 2015; Luger et al., 2019; see, e.g., Mandel & Agol, 2002). However this signature might be asymmetric due to several possible effects, one of which is the different temperature/pressure and/or chemical compositions the different terminator regions a transiting planet could have (see, e.g., Powell et al., 2019). Being able to model these asymmetric signatures directly from transit light curves could give us an unprecedented glimpse into planetary 3-dimensional structure, helping constrain models of atmospheric evolution, structure and composition.

`catwoman` is a Python package that models these asymmetric transit light curves, calculating light curves for any radially symmetric stellar limb darkening law and where planets are modelled as two semi-circles, of different radii, using the integration algorithm developed in (Kreidberg, 2015) and implemented in the `batman` library, from which `catwoman` builds upon. It is fast and efficient and open source with full documentation available to view at https://catwoman.readthedocs.io .

The light curves are modelled as follows: The decrease in flux, $\delta$, as a planet transits its star can be approximated by the sum

$$\delta = \sum_{i=1}^{N} I(x_m) \, \Delta A(x_m, R_{p,1}, R_{p,2}, \varphi, d), \qquad (1)$$

splitting the semi-circles into iso-intensity bands centred on the star and for each intersectional segment (see Figure 1) you multiply its area, $\Delta A$, by the intensity of the star and then sum these strips to generate the full $\delta$ for a specific separation between the centre of the star and planet, $d$. The code then increments $d$ by a small pre-determined amount (based on the time array given by the user) and recalculates $\delta$.

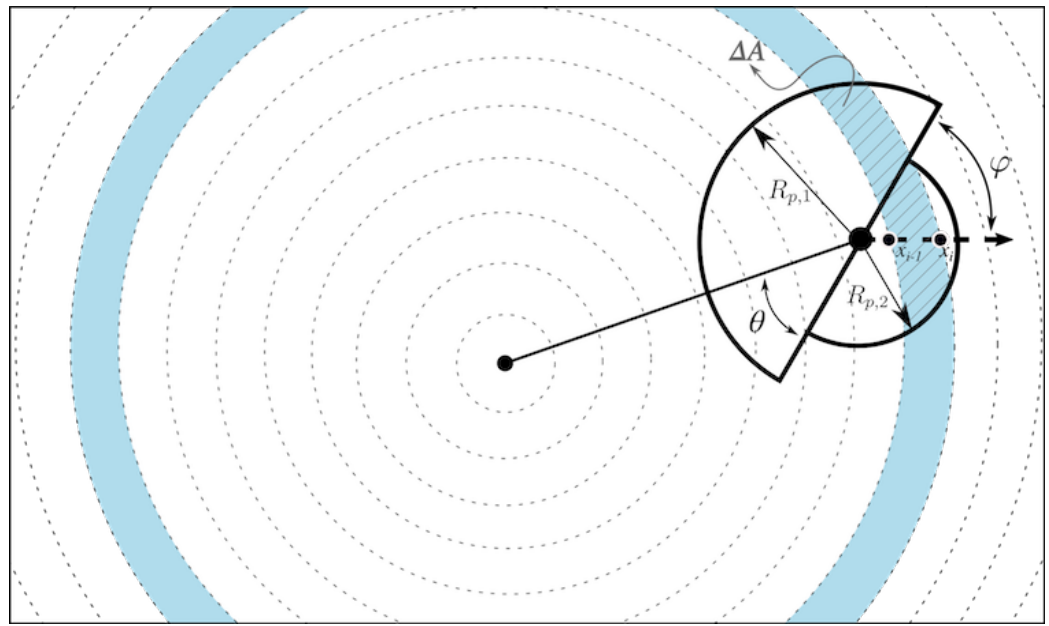

**Figure 1:** Diagram of the geometric configuration during transit of two stacked semi-circles (one of radius $R_{p,1}$, and another of radius $R_{p,2}$) that model the different limbs of an exoplanet transiting in front of a star. The area of the star has been divided in different sections of radius $x_i$ (dashed circles) — between each subsequent section, the star is assumed to have a radially symmetric intensity profile (e.g., blue band between $x_{i-1}$ and $x_i$ above). In order to obtain the light curve, the challenge is to calculate the sum of the intersectional areas between a given iso-intensity band and the semi-circles, $\Delta A$ (blue band with dashed grey lines). Note the stacked semi-circles are inclined by an angle $\varphi$ with respect to the planetary orbital motion.

The width of the iso-intensity bands determines the truncation error of the model. The model is first initialised with parameters including a maximum truncation error either set by the user or taken as the pre-set value as 1ppm. As in `batman`, `catwoman` first calculates many models, with varying widths and geometrically searches for a width that produces an error less than 1% away (and always less than) the specified level. The model then uses this width value to calculate the desired light curves. A lower specified error, and therefore thinner iso-intensity bands, produces more accurate light curves, however more steps are needed to calculate $\delta$ which takes more time.

`catwoman` also allows for $\varphi$, the angle of rotation of the semi-circles, to vary as a free parameter, which is something no other model has tried to implement, accounting for the possibility of spin-orbit misalignments of the planet. The two semi-circle radii, $R_{p,1}$ and $R_{p,2}$, and other orbital variables are also completely free parameters.

`catwoman` was designed to be used by astronomical researchers. For a realistic light curve with 100 in-transit data points, `catwoman` takes around 340 seconds to produce 1 million quadratic-limb-darkened light curves on a single 1.3 GHz Intel Core i5 processor. It is used in Espinoza & Jones (in prep.).

# Acknowledgements

We would like to thank the Max Plank Institute of Astronomy, Heidelberg, for providing the funding for this project and hosting Kathryn Jones as a summer student at the Institute.

# References

Kreidberg, L. (2015). batman: BAsic Transit Model cAlculatioN in Python. *Publications of the ASP*, *127*(957), 1161. https://doi.org/10.1086/683602

Luger, R., Agol, E., Foreman-Mackey, D., Fleming, D. P., Lustig-Yaeger, J., & Deitrick, R. (2019). starry: Analytic Occultation Light Curves. *Astronomical Journal*, *157*(2), 64. https://doi.org/10.3847/1538-3881/aae8e5

Mandel, K., & Agol, E. (2002). Analytic Light Curves for Planetary Transit Searches. *Astrophysical Journal, Letters*, *580*(2), L171–L175. https://doi.org/10.1086/345520

Powell, D., Louden, T., Kreidberg, L., Zhang, X., Gao, P., & Parmentier, V. (2019). Transit Signatures of Inhomogeneous Clouds on Hot Jupiters: Insights from Microphysical Cloud Modeling. *Astrophysical Journal*, *887*(2), 170. https://doi.org/10.3847/1538-4357/ab55d9